\documentclass[aps,twocolumn,amsmath,groupedaddress,10pt,prstab,floatfix]{revtex4-1}

\usepackage{adjustbox}
\usepackage{siunitx} % display numbers with units
\usepackage{graphicx} % recommanded by revtex for pictures
\usepackage{hyperref}

\begin{document}
\title{Continuous bunch-by-bunch spectroscopic investigation of the micro-bunching instability}

\author{Johannes L. Steinmann}
\email[]{johannes.steinmann@kit.edu}
%\thanks{}
\author{Tobias Boltz}
\author{Miriam Brosi}
\author{Erik Br{\"u}ndermann}
\author{Michele Caselle}
\author{Benjamin Kehrer}
\author{Lorenzo Rota}
\thanks{now at SLAC, Menlo Park, CA, USA}
\author{Patrik Sch{\"o}nfeldt}
\thanks{now at DLR-VE, Oldenburg, Germany}
\author{Marcel Schuh}
\author{Michael Siegel}
\author{Marc Weber}
\author{Anke-Susanne M{\"u}ller}
\affiliation{Karlsruhe Institute of Technology, Karlsruhe, Germany}

\date{\today}

\begin{abstract}
Electron accelerators and synchrotrons can be operated to provide short emission pulses due to longitudinally compressed or sub-structured electron bunches. Above a threshold current, the high charge density leads to the micro-bunching instability and the formation of sub-structures on the bunch shape. These time-varying sub-structures on bunches of picoseconds-long duration lead to bursts of coherent synchrotron radiation in the terahertz frequency range. Therefore, the spectral information in this range contains valuable information about the bunch length, shape and sub-structures. Based on the \textit{KAPTURE} readout system, a 4-channel single-shot THz spectrometer capable of recording 500 million spectra per second and streaming readout is presented. First measurements of time-resolved spectra are compared to simulation results of the \textit{Inovesa} Vlasov-Fokker-Planck solver. The presented results lead to a better understanding of the bursting dynamics especially above the micro-bunching instability threshold.
\end{abstract}

%\keywords{}

\maketitle

% ==============================
% PART 1: Introduction
% ==============================

\section{Introduction}
The self-interaction of a bunch with its own electric field can lead to potential well distortion and therefore to a deformation in the longitudinal phase space \cite{PhysRevSTAB.8.014202}. Above the micro-bunching instability (MBI) threshold, it results in the formation of sub-structures on the longitudinal bunch profile and increased emission of coherent synchrotron radiation (CSR) at wavelengths corresponding to the size of the sub-structures~\cite{PhysRevLett.90.094801}. The Karlsruhe Research Accelerator (KARA) is the electron storage ring of the test facility and synchrotron radiation source ANKA at the Karls\-ruhe Institute of Technology (KIT). In the short-bunch operation mode, this storage ring provides picoseconds-long bunches, which results in the emission of coherent radiation in the range up to a few terahertz (THz)~\cite{ICFA2012}.

Although the micro-bunching instability was predicted, simulated, and experimentally verified already in 2002, the bunch dynamics above the threshold are still not completely understood \cite{PhysRevSTAB.5.054402,PhysRevLett.89.224802,PhysRevLett.89.224801}. The outbursts of THz radiation have been observed at various facilities, i.e. 
ALS~\cite{PhysRevLett.89.224801}, %
ANKA~\cite{ANKA1}, %
Bessy II~\cite{Bessy1,PhysRevLett.88.254801,PhysRevLett.90.094801}, %
CLS~\cite{PhysRevAccelBeams.19.020704}, %
Diamond~\cite{Diamond1,Diamond2}, %
Elettra~\cite{Elettra}, %
MAX I~\cite{Max1}, %
MIT-Bates~\cite{PhysRevLett.96.064801}, %
MLS~\cite{PhysRevSTAB.14.030705}, %
NSLS VUV~\cite{NSLS_VUV1,NSLS_VUV2}, %
Soleil~\cite{Soleil}, %
SURF II~\cite{SURF} and III~\cite{PhysRevSTAB.4.054401}, %
UVSOR-II~\cite{UVSOR}%
.
The spectrum of the coherent emission depends on the bunch shape and therefore, the sub-structures created by the MBI. Observing the spectrum of the radiation emitted by the circulating electron bunches in the storage ring with single-shot detectors and turn-by-turn precision gives insights to the evolution of these structures. This is a complementary method to a direct observation of the electric field via electro-optical methods, providing a simpler experimental setup with faster repetition times~\cite{eosd,eosd3,Funkner:2018fmu,PhysRevLett.113.094801,PhysRevLett.118.054801}. Previous work by \textit{Finn et al.}~\cite{Finn} uses spatial segregation to record different frequencies simultaneously. In their current readout setup, the power and frequency components are analyzed in the frequency domain, dropping the phase relation between the different frequencies. In our setup, we use wire grids as broadband beam splitters and a KIT in-house developed readout system, providing synchronized data of the pulse amplitudes of each detector. Additionally, this
\textit{KAPTURE} readout system~\cite{KAPTUREa,KAPTUREb,KAPTUREc} provides continuous streaming of bunch-by-bunch, turn-by-turn, and single-shot data. 

One major step towards understanding the dynamics in micro-bunching instability would be the reconstruction of the longitudinal phase space density from measured data. A first step in this direction is the reconstruction of the longitudinal bunch profile. However, the used THz detectors only measure the emitted power of the synchrotron radiation while the phase information is lost. Unfortunately, a unique reconstruction of a signal only from the magnitude of its Fourier transform is mathematically impossible~\cite{Akutowicz}. Workarounds like a Kramers-Kronig dispersion relation give impressive results for known bunch shapes like a single Gaussian or cosine half wave~\cite{Schmidt:401413}. However, they fail for spikes and, especially, many small sub-structures as they happen for bunch charges above the MBI threshold being investigated here. Therefore, we chose the approach to use the Vlasov-Fokker-Planck solver \textit{Inovesa}~\cite{Inovesa} to simulate the phase space, extract the emitted power spectrum and compare this to our spectroscopic measurements. A comparison of the four spectral bands gives much more evidence about the validity of the simulation than the previous comparison to a single detector which integrates all the emitted radiation. A match between the simulation and all the detectors is a good indication that the simulated phase space represents the real bunch dynamics.

In this paper we first present general simulations by \textit{Inovesa}, showing the expected spectral emission during the MBI. Then, we introduce our 4-channel spectrometer setup. Good agreement between simulations and measurements is demonstrated at two chosen bunch currents exemplarily.

% ==============================
% PART 2: Inovesa Simulations
% ==============================

\section{Vlasov-Fokker-Planck Simulation}
Particle tracking including CSR effects is suitable for a single-pass machine, but still a computationally challenging problem for storage rings~\cite{PhysRevSTAB.11.030701}. As an alternative, the Vlasov-Fokker-Planck (VFP) equation can be used to simulate the longitudinal phase space density under the influence of a beam impedance~\cite{PhysRevSTAB.8.014202,PhysRevSTAB.14.061002}.
For that purpose, we use \textit{Inovesa}~\cite{Inovesa,inovesa:git}, an open source parallelized VFP-solver, developed at KIT. The simulations presented in the following use the CSR impedance with shielding by parallel plates~\cite{mkg-wakefield,warnock}. The parallel plates impedance model assumes an isomagnetic ring, where the electrons travel between two infinitely long and perfectly conducting parallel plates. Consequently, all interactions of the bunch with its emitted field in the straight sections are neglected in this model, as are possible interactions with the side walls.

Projecting the simulated longitudinal phase space onto the time axis provides the bunch profile. From that, the form factor and subsequently the emitted coherent radiation is calculated. Note that this is the bunch's radiation due to the parallel plates' impedance, which is not identical to the radiation observed at a beamline.  The calculation of the radiation measured at the beamline would require the inclusion of additional effects such as beamline apertures, which attenuate lower frequencies, and transmission properties of the used vacuum windows and mirrors. However, for the comparison in this paper, we omitted these effects as well as source specific effects like edge radiation and polarization.

In our simulations, we used the bending radius (\SI{5.559}{m}) and beam pipe height (\SI{32}{mm}) of KARA. All ``non-static'' properties, like synchrotron frequency and RF voltage, have been chosen to fit to the measurements presented in the following (see Table~\ref{table:measurement}).

\begin{figure}[htb]
  \centering
   \includegraphics[width=\columnwidth]{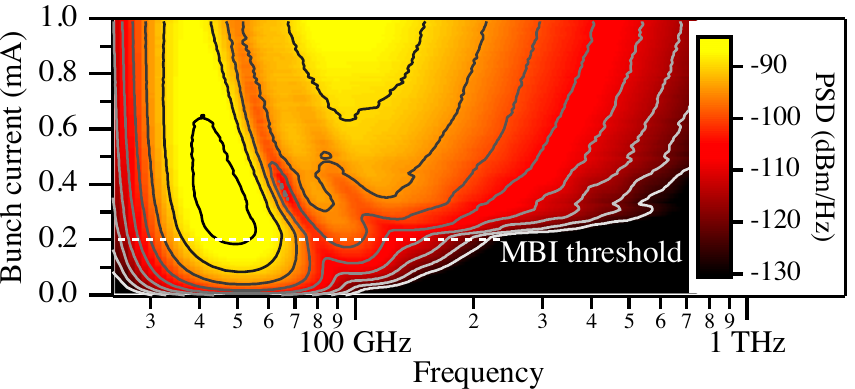}
   \caption{\textit{Inovesa} simulation: Average emitted power spectral density (PSD) for different bunch currents. The contour lines are in \SI{5}{dB} distance between \SI{-85}{dBm} and \SI{-130}{dBm}.}
   \label{fig:sim_avg_spectrogram}
\end{figure}

The simulated spectrogram in Fig.~\ref{fig:sim_avg_spectrogram} shows the average emitted CSR spectrum at different bunch currents. In the zero current limit, the bunch has a minimum bunch length and a Gaussian shape without any sub-structures. With increasing current, the bunch form changes due to potential-well distortion. This leads to a decrease of the intensity of the first frequency peak around \SI{50}{GHz} when compared to the quadratic current-dependency expected for constant bunch shapes. On the other hand, the distortion of the Gaussian shape introduces spectral components at frequencies above \SI{100}{GHz}. At bunch currents above the MBI threshold, additional sub-structures form and even higher frequency components arise. At the same time, bunch lengthening brings a decreasing height of the first peak. The threshold current in this simulation is located between \SIrange{0.22}{0.23}{\mA}, while the simulated current steps were \SI{10}{\micro\ampere}.

\begin{figure}[b]
  \centering
   \includegraphics[width=\columnwidth]{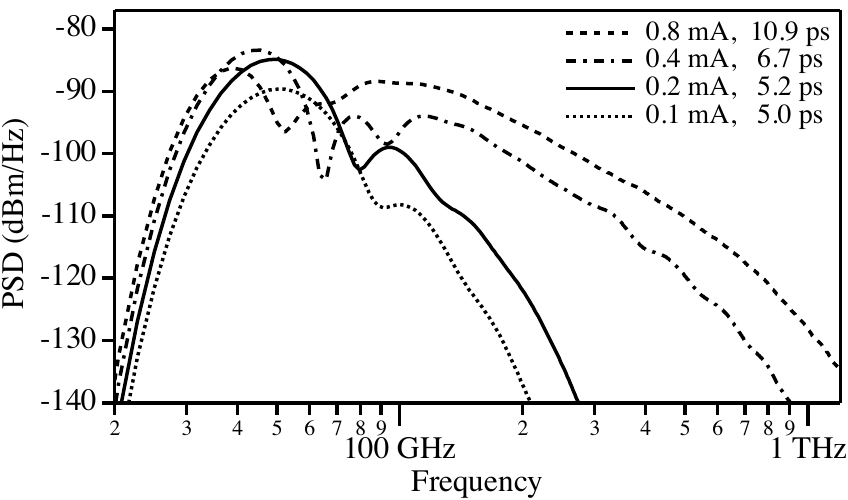}
   \caption{\textit{Inovesa} simulation: Slices for selected bunch currents and average rms bunch lengths of the emitted CSR spectrum in Fig.~\ref{fig:sim_avg_spectrogram}.}
   \label{fig:sim_avg_spectrum}
\end{figure}

Figure~\ref{fig:sim_avg_spectrum} shows in greater detail four specific averaged spectra at \SI{0.1}{\mA}, \SI{0.2}{\mA}, \SI{0.4}{\mA} and \SI{0.8}{\mA} with an average rms bunch length of \SI{5.0}{ps}, \SI{5.2}{ps}, \SI{6.7}{ps} and \SI{10.9}{ps}, respectively. 

\begin{figure*}
    \includegraphics{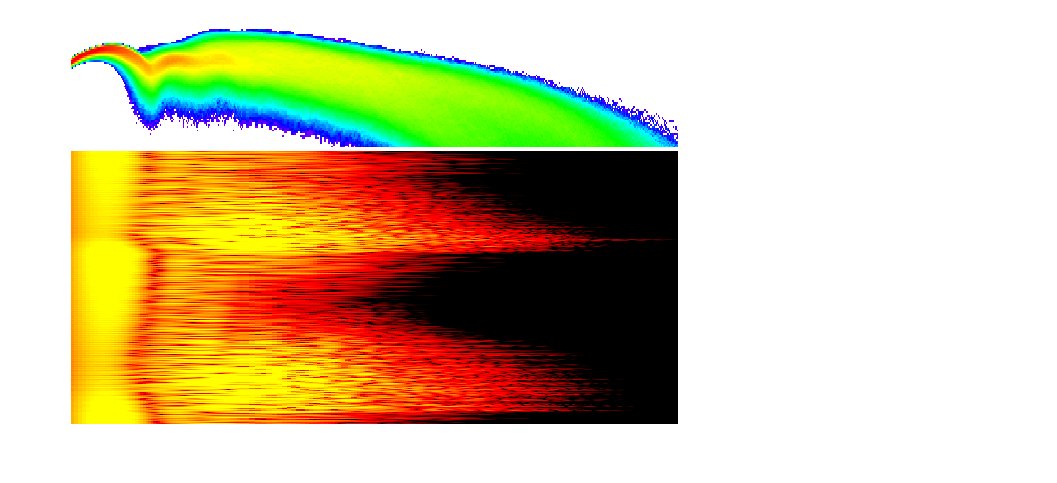}%
    \llap{{%  move next graphics to top right corner
      \includegraphics{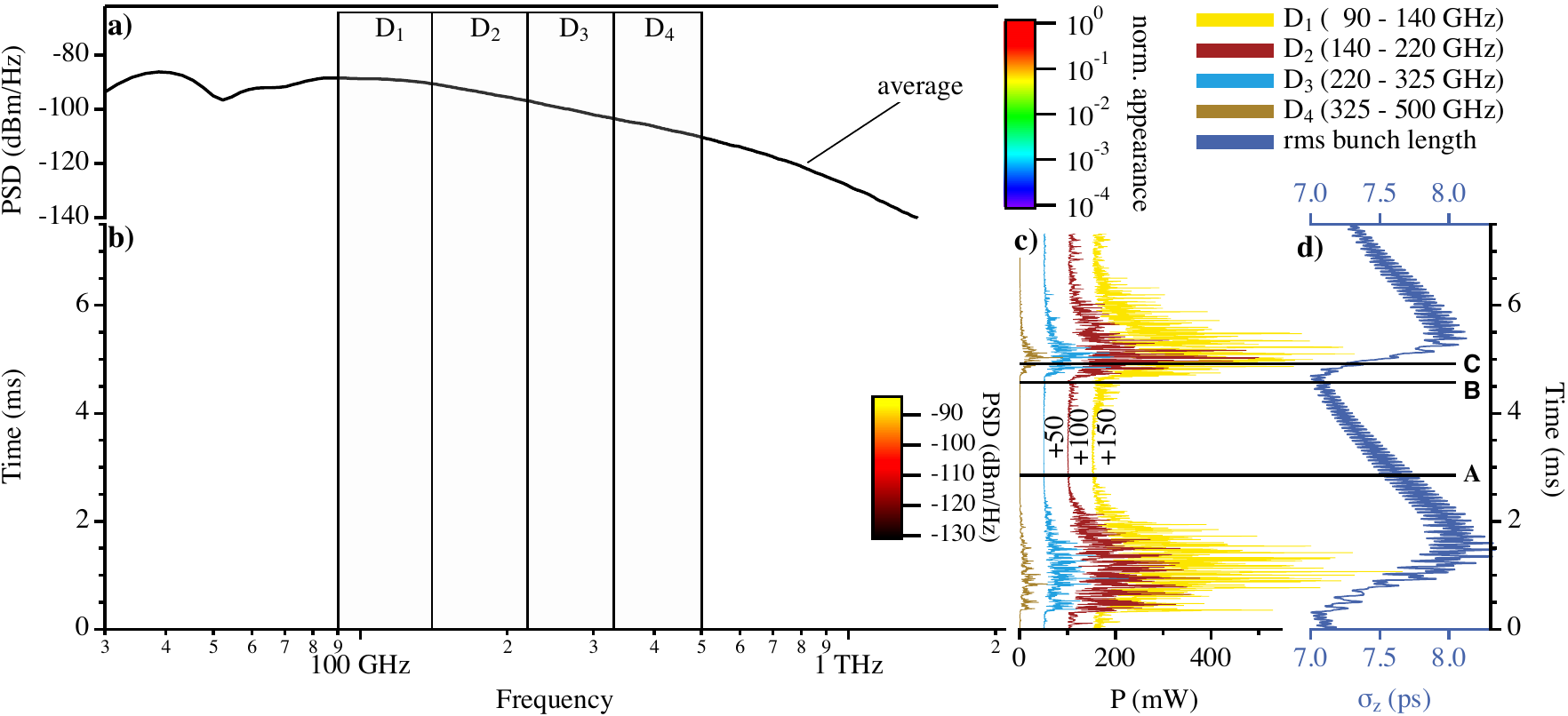}%
    }}
   \caption{\textit{Inovesa} simulation: Bursts at \SI{0.8}{m\ampere}: a) shows the average emitted spectrum on top of a histogram displaying the normalized appearance of frequency and power spectral density (PSD) values. For more information see text. The image in b) shows the instantaneous emitted coherent radiation due to the changing bunch profile over time. The expected integrated power P in the marked frequency bands of the used Schottky diodes are shown in c) and the corresponding rms bunch length $\sigma_z$ in d) for the same time axis. For better visibility, the signals of $D_1$, $D_2$ and $D_3$, are shifted by 150, 100 and \SI{50}{mW}, respectively. For selected times A, B and C see details in Fig.~\ref{fig:SimSpectra}.}
   \label{fig:InovesaTimeresolved}
\end{figure*}

On the short timescales, the MBI leads to different behaviours depending on the machine settings and the bunch current. For all machine settings which we have analyzed so far, multiple regimes can be categorized. Two of them will be analyzed in more detail in the following. On the one hand, the region just above the instability threshold. It features regular modulations and occurrence of sub-structures and is called \textit{regular bursting regime}. On the other hand, a regime with a sawtooth behaviour where periodic outbursts of radiation and rising sub-structures are followed by a decay and stabilization phase. In the machine settings presented here, this \textit{sawtooth bursting regime} lies above three times the threshold current.

At this high bunch current, the formed sub-structures are self-amplified by their wake potential, quickly driving the instability further. Consequently, the energy spread and the bunch size grows. Hereafter, the emitted coherent radiation and, thus, the additional driving wake potential vanishes due to the decreased charge density. Radiation damping smooths the bunch shape, too. Subsequently, the bunch shrinks according to the damping until the MBI threshold is reached again. The average emitted power spectrum is therefore dependent on the duration, intensity and repetition rate of the outbursts while below the threshold it is governed by static bunch deformation.

Figure~\ref{fig:InovesaTimeresolved} presents simulated data for a bunch current of \SI{0.8}{\mA}, more than three times the threshold current. The changing instantaneous CSR spectrum over time is calculated from the simulated bunch profiles. For comparison to slower data acquisition systems and spectrometers, the average spectrum is displayed together with the normalized appearance of all spectra, which are emitted within \SI{50}{\ms}. The normalized appearance has been calculated by a histogram of every frequency bin, therefore counting the emitted intensities at every frequency over time. Then it was normalized to the number of analyzed spectra. 

Most of the emitted spectra are below the average (note the logarithmic scale), because the high frequencies and high intensities are emitted only for a short time during a burst. The fact that the coherent enhancement increases the emitted power by many orders of magnitude, leads to an average above the median spectrum. Also indicated are four frequency bands (Fig.~\ref{fig:InovesaTimeresolved}): $D_1$ (\SIrange{90}{140}{GHz}), $D_2$ (\SIrange{140}{220}{GHz}), $D_3$ (\SIrange{220}{325}{GHz}) and $D_4$ (\SIrange{325}{500}{GHz}). The same frequency bands are used in the described measurements (see Table~\ref{table:vdi_schottky}).

Figure~\ref{fig:InovesaTimeresolved} additionally shows the emitted THz intensity in these four spectral bands and the rms bunch length as a function of time. The small, but fast, bunch-length modulation between the radiation outbursts can be modelled by a quadrupole motion, which leads to a periodicity with twice the synchrotron frequency. On the longer timescales, the rms bunch length decreases due to radiation damping, until the threshold is reached. Then, the strong micro-bunching instability drives sub-structures and blows up the bunch. This blow-up in combination with diffusion and damping in phase space homogenizes the bunch shape and the sub-structures vanish. Accordingly, the bunch length and the energy spread fluctuation show the same periodicity as the bursting behavior (see also~\cite{kehrer}). Because the emitted spectrum is sub-structure dominated and only the spectral intensity is measured (i.e. the phase is lost), a direct reconstruction of the bunch  profile is not possible.

\begin{figure}[hbt]
  \centering
   \includegraphics{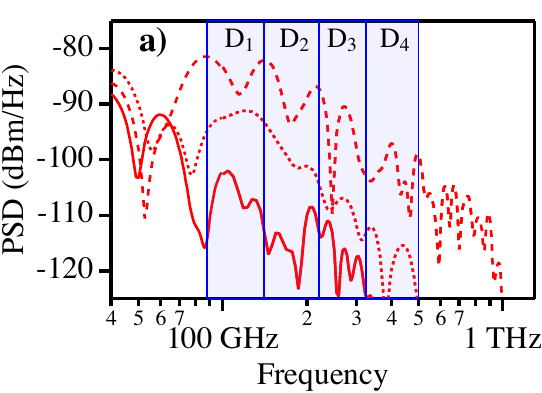}\hfill
   \includegraphics{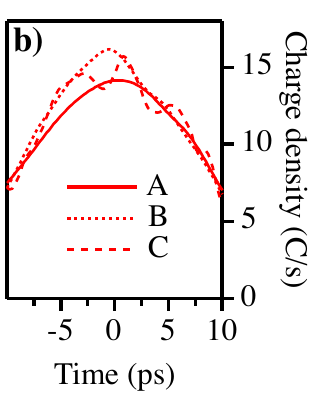}
   \caption{Calculated emitted spectrum a) and bunch profile b) at selected times A, B and C of Fig.~\ref{fig:InovesaTimeresolved} as a function of spectral frequency and time, respectively. The high frequencies are very unstable and only emitted for a short time during a radiation outburst. However, the intensities emitted by the sub-structures are many orders of magnitude higher due to the coherent enhancement. The indicated bands $D_1$, $D_2$, $D_3$ and $D_4$ (shaded blue) correspond to the detector sensitivity used in the measurements. For details see Fig.~\ref{fig:InovesaTimeresolved} and the text.}
   \label{fig:SimSpectra}
\end{figure}

Three points in time (A, B and C) are shown in more detail in Fig.~\ref{fig:SimSpectra} with their spectrum and bunch profile. Time A lies between two outbursts (cf.~Fig.~\ref{fig:InovesaTimeresolved}(d)): the shape is smooth since the sub-structures have already decayed, and the bunch is further shortening. At the high frequencies of $D_3$ and $D_4$, very low power is emitted. At time B, the shortened bunch implies an increased wake potential, which leads to a bunch deformation, visible in the bunch profile (cf.~Fig.~\ref{fig:SimSpectra}(b), trace B) and drives the micro-bunching instability. This results in more radiation, seen first at lower frequencies (cf.~Fig.~\ref{fig:SimSpectra}(a)), and also increases the intensity of the wake potential. At time C, the sub-structures, now clearly visible on the bunch profile, have the highest amplitude and shortest size leading to the outburst of radiation as seen at all frequency bands. The high wake potential blows up the bunch quickly, which, due to diffusion and damping, becomes stable again and shrinks until the next burst occurs. The described process takes a few milliseconds, only.

% ==============================
% PART 3: Measurement Setup
% ==============================

\section{Measurement Setup}
Measurements have been performed at the storage ring KARA at KIT and the radiation was coupled to the detection system via the infrared beamline IR2~\cite{IR}. The storage ring was operated in a short-pulse mode (low-alpha operation), wehere the magnet optics were set to obtain a momentum compaction factor $\alpha_c$ of \num{5e-4}. With the \SI{800}{kV} accelerating voltage, this results in a calculated zero-current bunch length of \SI{4.5}{ps}. Initially, the bunch current of the observed bunch was \SI{2}{mA}, which is way above the MBI threshold~\cite{ICFA2012}, and decayed to \SI{0.18}{mA}, which is below the threshold. The scaling law~\cite{PhysRevSTAB.13.104402} for the instability predicts for this machine settings a MBI threshold current of \SI{0.2}{\mA}. The important machine parameters of the measurement are summarized in Table~\ref{table:measurement}.

\begin{table}[bht]
\begin{ruledtabular}
  \centering
  \caption{\label{table:measurement}Machine paramenters during the experiment}
  \begin{tabular}{lrl}
	Beam energy, \cite{IPAC15Chang} & 1.287 & GeV\\
	Circumference &	 110.4 & m \\
	Vacuum chamber full height & 32 & mm\\
	Revolution frequency &	2.7157 & MHz \\
	RF frequency & 499.71 & MHz \\
	RF amplitude & 799.2 & kV \\
	Synchrotron frequency & 8.2 & kHz \\
	Estimated MBI threshold, \cite{PhysRevSTAB.13.104402} & 0.2 & mA\\
	Calc. momentum compaction $\alpha_c$ & \num{5e-4}& \\
	Calc. relative energy spread & \num{0.47e-3} & \\
	Calc. zero-current bunch length (rms) & 4.5 & ps\\
  \end{tabular}
  \end{ruledtabular}
\end{table}

\begin{figure}[b]
  \centering
   \includegraphics[width=.45\columnwidth]{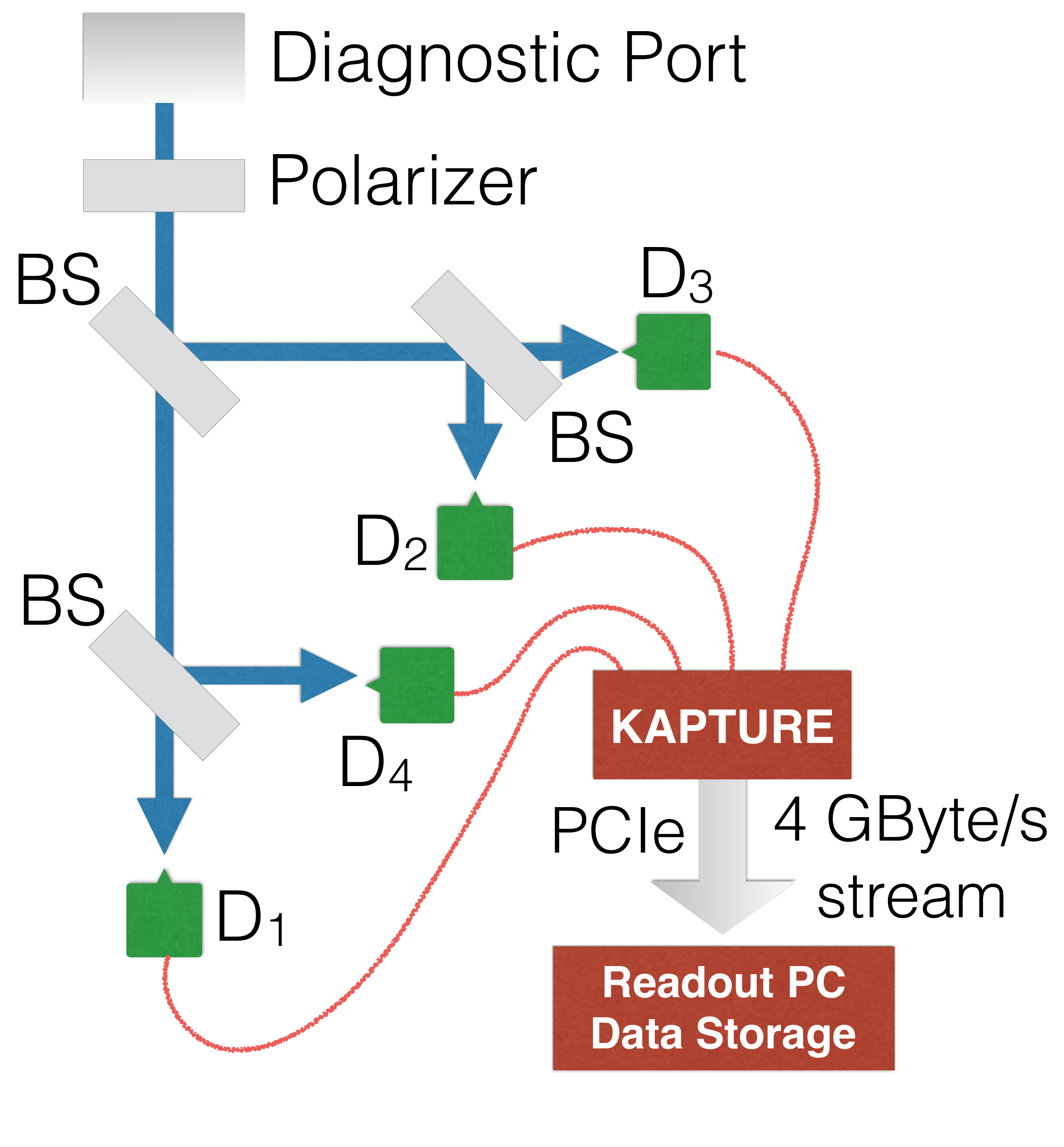}\includegraphics[width=.55\columnwidth]{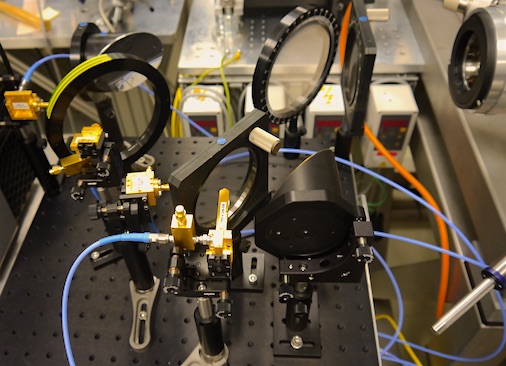}
   \caption{The synchrotron radiation emitted at the diagnostic port is first polarized and then divided into four beams by three wire-grid beam splitter (BS). The split beams are focused onto four detectors, each one sensitive in its specific frequency band (see Table~\ref{table:vdi_schottky} and  Fig.~\ref{fig:InovesaTimeresolved} for details). The single-shot measurements are sampled and read out by the \mbox{\textit{KAPTURE}} system.}
   \label{fig:setup}
\end{figure}

The setup at the beamline is sketched in Fig.~\ref{fig:setup}. The radiation was coupled out through a Z-cut quartz window at the diagnostic port of the infrared beamline. The optics of the beamline were designed to project an image of the entrance edge of the radiating bending magnet. However, the long wavelengths suffer from diffraction and limiting apertures. This was confirmed by Fourier transform infrared (FTIR) spectroscopy at the beamline measuring the maximum power around \SI{200}{GHz}~\cite{IPAC2014} with strong suppression of frequencies below \SI{100}{GHz}.

To ensure a purely linear polarization of the incoming CSR, the beam is pre-polarized and afterwards, with a set of three additional wire-grid polarizers, split into four partial beams of nearly equal power, each focused to a commercially available wave guide coupled Schottky barrier diode (SBD) detector~\cite{VDI}. Each detector is sensitive in a different frequency band, spanning the range from \SIrange{90}{500}{GHz} (Tab.~\ref{table:vdi_schottky}). The detector signals were simultaneously read out with \mbox{\textit{KAPTURE}}'s four input channels~\cite{KAPTUREa,KAPTUREb,KAPTUREc}.

\mbox{\textit{KAPTURE}} is synchronized to the 500-MHz master clock of the storage ring and digitizes the detector signals with the same repetition rate. The sampling time can be set with a \SI{3}{ps} accuracy for each input channel, individually. A track-and-hold circuit in front of the analog-to-digital converter (ADC) stores the signal amplitude at the chosen time until it is read out by the 12-bit ADC. The results presented here are based on data sets recorded in intervals of 10 seconds to reduce the amount of stored data. Each data set consists of \SI{100}{ms} continuous turn-by-turn data for all 184 RF buckets, i.e. every bunch has continuously been tracked for more than \num{2.7e5} consecutive turns.

Table~\ref{table:vdi_schottky} lists the different frequency bands, the diode models, and the average DC responsivity of the used detectors.
\begin{table}[thbp]
\caption{\label{table:vdi_schottky}Schottky barrier diode detectors used~\cite{VDI}}
\begin{ruledtabular}
    \begin{tabular}{lccc}
    Label   &	Band (GHz)  &   VDI Model   & DC resp. (V/W)\\\colrule
    $D_1$   &	90-140      &   WR8.0ZBD    &	2000\\
    $D_2$   &	140-220     &   WR5.1ZBD    &	2000\\
    $D_3$   &	220-325     &   WR3.4ZBD    &   1500\\
    $D_4$   &	325-500     &   WR2.2ZBD    &   1250\\
    \end{tabular}
  \end{ruledtabular}
\end{table}

The responsivity shown is only indicative of the sensitivity of the complete detector system, i.e. the diodes should not be compared quantitatively, because the measured synchrotron radiation pulse is broad-band and the frequency acceptance of the diodes is not uniform over the entire band. To improve the used ADC range, three of the SBD detectors ($D_2$, $D_3$ and $D_4$) were amplified using a \SI{15}{dB} amplifier with \SI{18}{GHz} bandwidth.

Even though the analog frequency bandwidth of all diodes has been measured to be above \SI{18}{\GHz}, the synchrotron emits ps and sub-ps pulses shorter than the overall response time of the SBD detector circuit. Therefore, the measured pulse amplitude is dominated by the impulse response due to the quality of the RF-readout path of each individual diode. In comparison to the individual DC-responsivities, the THz impulse response is very difficult to calibrate. Thus, to not bias the reader, the data shown here was not corrected for the diode's DC responsivities, however, we don't assume these properties to change during the experiment. The reader has to keep in mind that quantitative comparisons between the diodes require special care.

If the bunch shape would be known, this setup could be used as a single-shot bunch length monitor for bunches in the picosecond range. However, above the MBI threshold, the bunch shape is dominated by the sub-structures and the instantaneous overall bunch-length can not be determined.

% ==============================
% PART 4: Measurement Results
% ==============================

\section{Measurement Results}

\subsection{Instability Threshold Determination}
\begin{figure}[tb]
  \centering
   \includegraphics[width=\columnwidth]{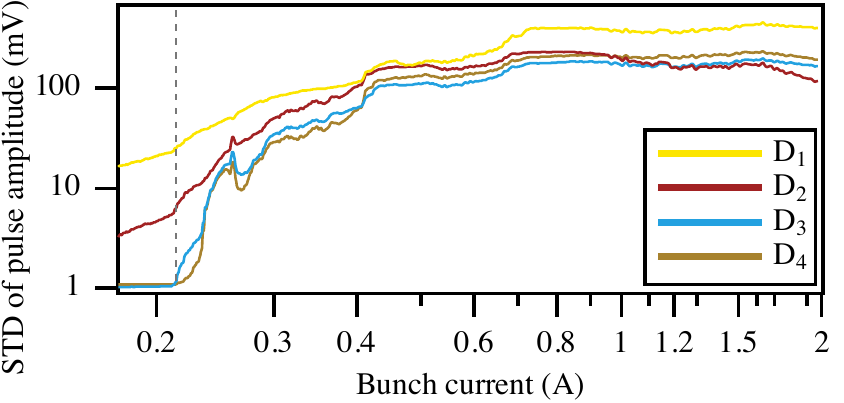}
   \caption{Measured standard deviation (STD) of the pulse intensity in the different diode frequency bands. Above the determined MBI threshold of \SI{0.214(2)}{\milli\ampere} (vertical dashed line), the fluctuations increase significantly.}
   \label{fig:pulse_amplitude_STD}
\end{figure}

An important parameter to determine is the MBI threshold current at the given machine setting. A good method for its determination is to measure the fluctuations of the THz signal amplitude~\cite{PhysRevAccelBeams.19.110701}. For that reason, the standard deviation of the detector amplitude is analyzed as shown in Fig.~\ref{fig:pulse_amplitude_STD}.

Below the threshold current, only two diodes ($D_1$, $D_2$) detect fluctuations in the pulse intensity, because no CSR is emitted at the higher frequencies ($D_3$, $D_4$, cf. simulation of the emitted spectrum in Fig.~\ref{fig:sim_avg_spectrogram}). The fluctuations below the threshold are in accordance with a quadrupole motion model of the synchrotron oscillation that slightly modulates the bunch length. The slope increases significantly at the threshold current, where additional sub-structures influence the emitted radiation.

The determined MBI threshold current of the measurement is at \SI{0.214(2)}{\milli\ampere}, which is in agreement to the threshold predicted by the scaling law (\SI{0.20(1)}{\milli\ampere}) and the one simulated by \textit{Inovesa} (\SI{0.225(5)}{\milli\ampere}). Uncertainties on the measured input parameters account for the error in scaling law value but have not been considered in the VFP simulation. 

Above the threshold, the standard deviation increases until it remains at a high level, which is even above the average amplitude. That is due to the outbursts of radiation followed by relatively low power during the shrinking of the bunch. Consequently, long averaging times are needed in this regime, if one intends to do conventional spectroscopy with typically, in comparison to our measurements, slower data acquisition systems.

\begin{figure}
   \includegraphics{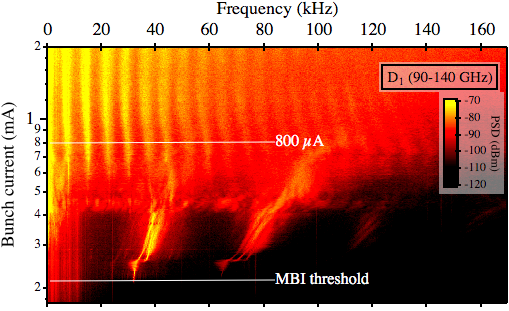}\\
   \includegraphics{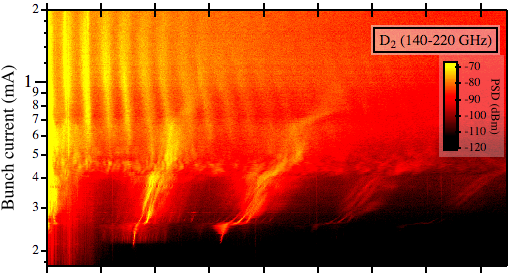}\\
   \includegraphics{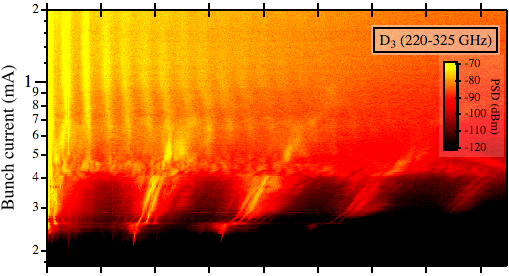}\\
   \includegraphics{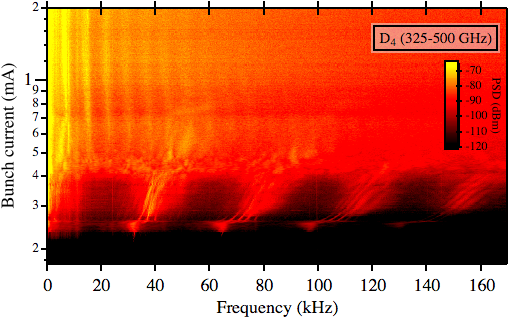}
   \caption{Bursting spectrograms of each detector. The main features are visible in all spectrograms, albeit with different contrast. The color scale is a logarithmic scale of the detector output power into \SI{50}{Ohms} and should not be compared quantitatively between the diodes due to differing frequency bandwidth, responsivity and RF readout efficiency of the diodes. Also note, that the $D_1$ detector was operated without amplifier, whereas all other diodes were amplified by \SI{15}{dB}.}
   \label{fig:spectrogram_lin}
\end{figure}

\begin{figure*}[tb]
  \centering
   \includegraphics[width=\textwidth]{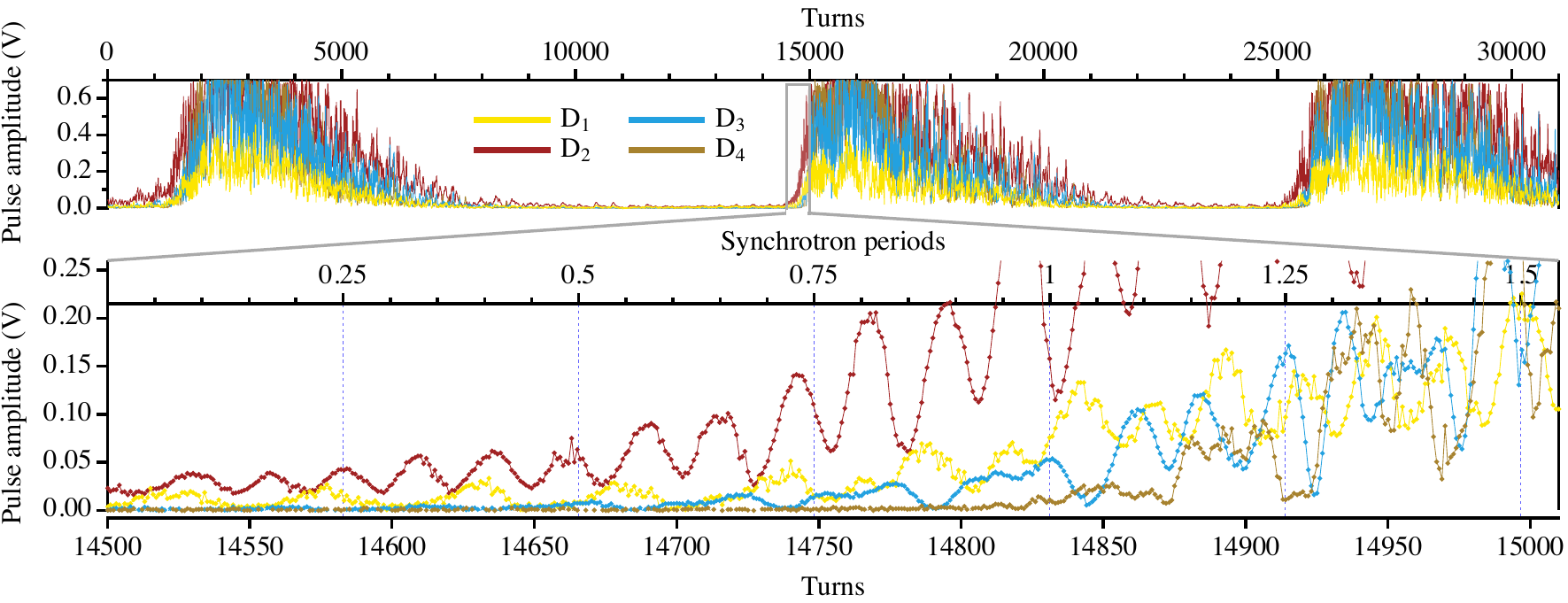}
   \caption{Measured turn-by-turn amplitudes of the four different diodes (1000 turns $\approx$ \SI{0.368}{\ms} $\approx$ 3 synchrotron periods). The bottom panel shows a zoom-in to the beginning of an outburst. Every point is a measured pulse amplitude, the connecting line is a guide to the eye.}
   \label{fig:timedomain}
\end{figure*}

\subsection{Bursting Spectrograms}

So-called \textit{bursting spectrograms} have proven to be an excellent tool to analyze the bursting dynamics~\cite{Kuske:2010ppa,Diamond2,Soleil,PhysRevSTAB.14.030705}. For this purpose, the time domain data of THz pulses and their time evolution is Fourier transformed. The resulting spectrum is plotted over each corresponding measured bunch current. For a more detailed description of the data analysis strategy, see~\cite{PhysRevAccelBeams.19.110701}. Such a spectrogram, showing specific modulation frequencies of the THz power, is reproducible and depends on machine parameters like the momentum compaction factor, the accelerating voltage and the impedance of the storage ring. It is a fingerprint of the accelerator in its current state. Figure~\ref{fig:spectrogram_lin} shows the spectrogram for each of the four detectors. Different current-dependent bursting regimes can be identified~\cite{PhysRevAccelBeams.19.110701}.

Above the MBI threshold at \SI{0.214(2)}{\milli\ampere}, the instabilities start with high frequency modulations around \SI{32}{kHz}. With increasing current, also low frequencies arise that indicate the occurrence of periodic bursts. The repetition rate of the bursts is below \SI{1}{kHz} and hardly visible in the plots. After an intermediate stage with more complex structures above \SI{0.7}{\milli\ampere}, the spectrum changes  to a regular pattern of harmonics near the synchrotron frequency. With increasing bunch current the amount of white noise also rises. Since the ADC pre-amplifier has not been changed, these are not digitization-induced fluctuation effects, but account for a high amount of randomness in the bursting behavior. As a consequence, even though the form of the bursts is similar and their appearance repeats, none of them are identical. 

Furthermore, it can be seen that the main features of the spectrogram are the same, although the different diodes span different frequency ranges from \SIrange{90}{500}{GHz}. Differences in the spectrogram can be seen in relative intensity variations of some features, the amount of white noise and the number of visible harmonic structures.

\subsection{Sawtooth Bursting Regime}

In the following, a time-domain dataset at \SI{0.8}{\milli\ampere} is analyzed in more detail. This current is about four times the MBI threshold and in the regime of large periodic bursts. Due to bunch lengthening, the average bunch length is rather long, which limits the amount of coherent radiation. Nevertheless, the strong instability with many small sub-structures in the electron density leads to periodic bursts of high amplitude THz radiation.

Turn-by-turn data of the four detectors are shown in Fig.~\ref{fig:timedomain}. In agreement with the simulations (compare to Fig.~\ref{fig:InovesaTimeresolved}), the pulse amplitudes of the $D_1$ and $D_2$ diodes are on an almost constant and low level before the burst. They are mainly modulated with twice the synchrotron frequency in accordance with the quadrupole motion model (not visible in the plots). When the burst starts, as seen in the zoom-in, the observed modulation changes, which is due to the forming sub-structures rotating in phase space. The diodes, sensitive at a higher frequency band, do not measure notable coherent radiation until the burst starts. Additionally, the burst's observation occurs later compared to the lower frequencies bands.

\subsection{Regular Bursting Regime}
The bursting regime slightly above the MBI threshold, we refer to as \textit{regular bursting}~\cite{SWB-486190137}. Here, besides the always visible synchrotron frequency, only a single modulation frequency (the \textit{regular bursting frequency}) and its harmonics are visible. The actual frequency depends on the number of structures in the phase space and on the synchrotron frequency~\cite{PhysRevSTAB.17.010701}.

\begin{figure}[hbt]
   \includegraphics{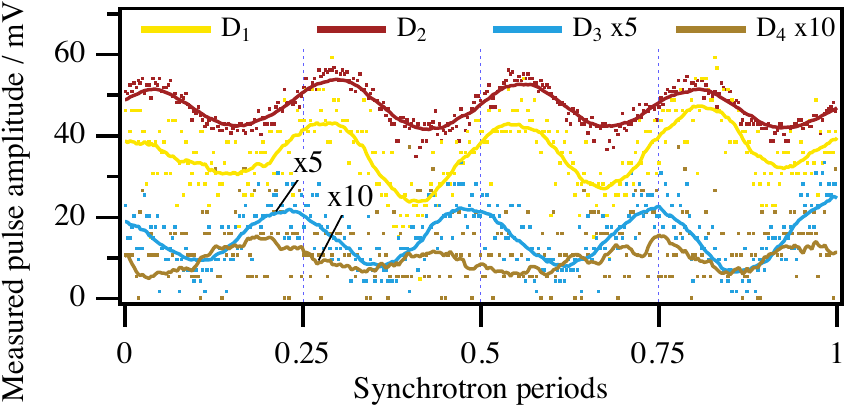}\\[8pt]
   \includegraphics{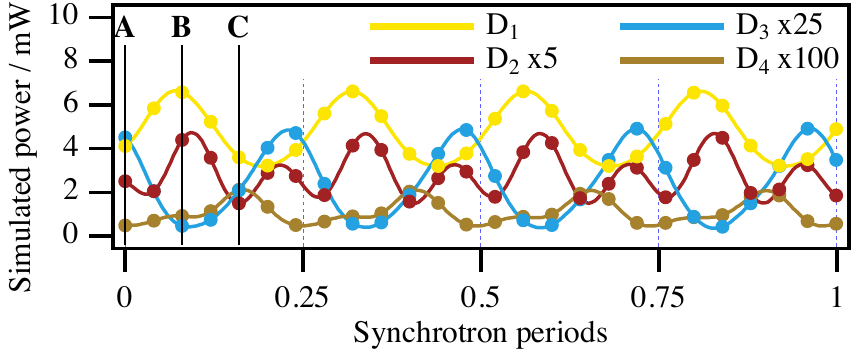}
   \caption{Measurement (upper) and simulation (lower) in the regular bursting regime. The pulse amplitudes measured every turn by KAPTURE are shown as dots while the solid lines are smoothed by a box car filter of 30 turns (\num{0.09} synchrotron periods). Both plots span a complete incoherent synchrotron oscillation. The visible quantization of the $D_4$ detector signal is due to the low signal strength at the high frequencies in combination with limited ADC resolution.
   Between the observed bands, a clear phase shift can be seen.
   The points marked in the lower plot are shown in more detail in Fig.~\ref{fig:fingerplot}.
   }
   \label{fig:finger}
\end{figure}

The upper part of Fig.~\ref{fig:finger} shows a measurement with the four detectors in that regime. Due to the low intensity at high frequencies, the signal of the $D_4$ detector is limited by the ADC resolution.
The same regular bursting frequency around \SI{32}{kHz} is observed in all diodes which was also observed in the spectrograms (cf. Fig.~\ref{fig:spectrogram_lin}). In addition to the spectrogram, which only shows the magnitude of the oscillation, a phase shift between the detectors can be seen in the time domain. This is not an artifact but due to the specific structure rotating in phase space. The projection of that structure, i.e. the bunch profile, enlarges and shrinks according to the phase space rotation. Therefore, a different intensity is observed in the distinct spectral ranges.
A similar behavior can be observed in the simulation. The simulated time domain data for the same machine settings is displayed in the lower panel of Fig.~\ref{fig:finger} . Since \textit{Inovesa} simulates the complete phase space, the source of the observed phase shift can be investigated.

\begin{figure*}
   \includegraphics[width=\textwidth]{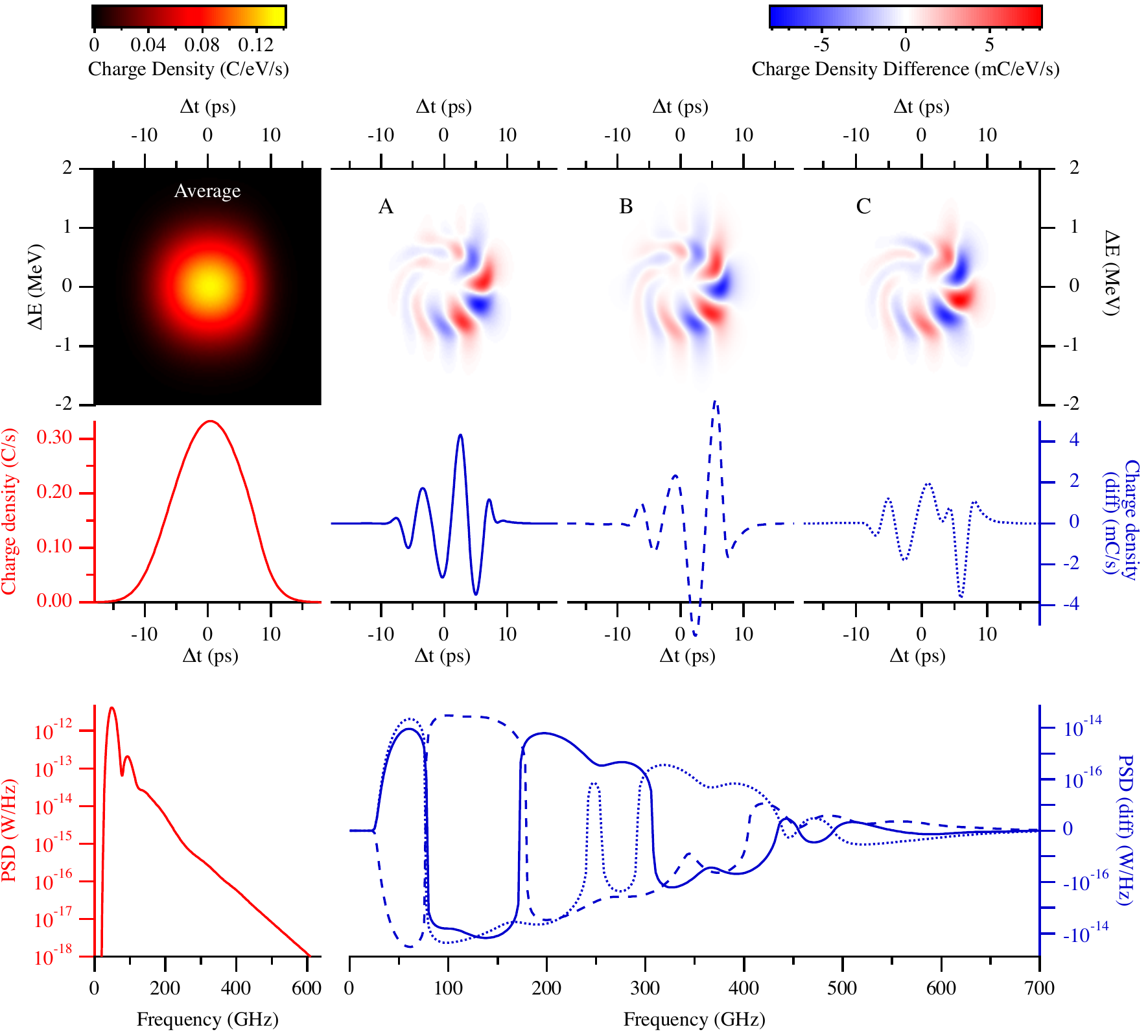}
   \caption{\textit{Inovesa} simulation: The three marked time steps of Fig.~\ref{fig:finger} are investigated in greater detail. The phase space as simulated by \textit{Inovesa} (top) is used to calculate the longitudinal charge density (middle) and from that the emitted spectrum shielded by parallel plates (bottom). The bunch head is to the right and the rotation in phase space is therefore counter-clockwise. For better visibility, the average was subtracted from all time steps. The first column shows the subtracted time-average phase space (first line), time-average charge density profile (second line), and average emitted radiation spectrum (third line).\\ 
   Note that the spectrum of the average profile is not the average spectrum, shown in the bottom line. In particular, the average profile does not feature any trace of the sub-structures, so its spectrum does not reach the same high frequencies. In the top line, the averaged phase space is subtracted and the sub-structures are therefore clearly visible, where the red color denotes a charge density higher than the average and the blue color denotes a density lower than the average. The three spectra have more or less power than the average, dependent on the individual bunch profile. To account for that and nevertheless present the differences over a wide range, a two sided logarithmic scale is used where differences below \SI{1e-18}{W/Hz} have been treated as the zero value.
   }
   \label{fig:fingerplot}
\end{figure*}

The three marked time steps of Fig.~\ref{fig:finger} are shown in more detail in Fig.~\ref{fig:fingerplot}. For better comparison, the average phase space is subtracted from the detailed views. The phase space at that current has permanent substructures that rotate counter-clockwise. \textit{Permanent} in this case means that the sub-structures are always present, and the phase space replicates with the observed regular bursting frequency. However, the structures themselves evolve and change in amplitude during their rotation (synchrotron motion) in phase space. At the bunch head, the sub-structures are largest, while they almost vanish on the opposite side. At these machine settings, five minima and maxima in the phase space are seen. The phase space image replicates with the regular bursting frequency, which is approximately four times the synchrotron frequency. The different impact on the spectrum at the three time points explain the different behavior of the narrow-band detectors. This effect can not be observed when using a single broadband detector that integrates the whole sensitivity range.

\subsection{Statistical Investigation}
\begin{figure}[bt]
  \centering
   \includegraphics{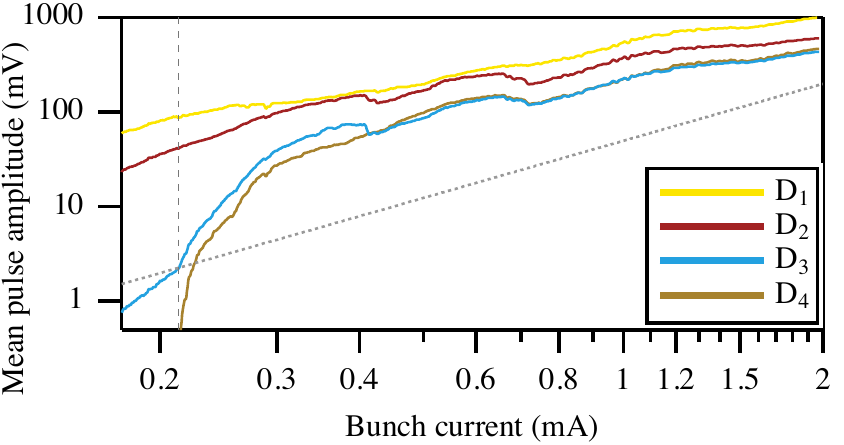}\\[10pt]
   \includegraphics{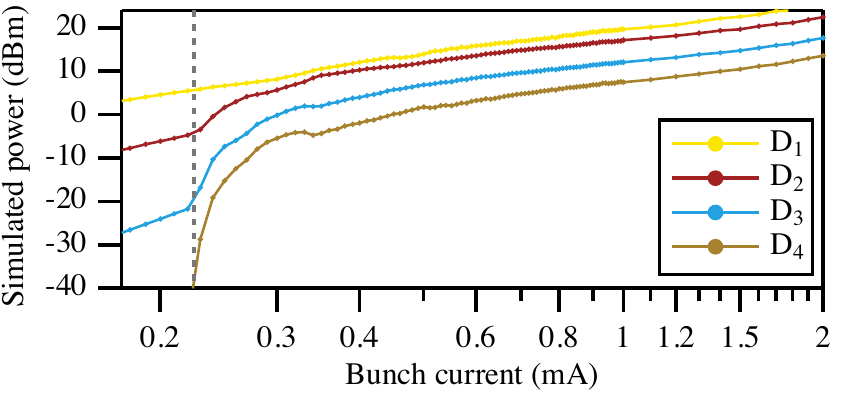}
   \caption{Upper: Average measured amplitude of the single pulse signal over the bunch current corresponding to the average optical power emitted in each frequency band. The $D_1$ data was compensated to account for the missing amplifier. For a stable bunch profile, a quadratic dependency would be expected. To guide the eye, the dotted line shows such a behaviour. The MBI threshold appears at \SI{0.214(2)}{mA} (vertical dashed line).
   Lower: Simulated, average integrated CSR power emitted in each frequency range of the measurement. The connecting line between the simulation steps is a guide to the eye. The vertical dashed line denotes the simulated MBI threshold at \SI{0.225(5)}{\milli\ampere}. This is the simulated power emitted by the bunch over the whole turn. For details see text.
   }
   \label{fig:pulse_amplitude_avg}
\end{figure}

Because of the uniqueness of the bursts, especially in the sawtooth regime at currents far above the instability threshold, a direct comparison between measurement and simulation is difficult. Even though a burst may look similar in simulation and experiment, its underlying physics do not necessarily have the same origin. To cross-check both, averaged values are compared derived from our simulations with \textit{Inovesa} and the measurements obtained with our experimental setup. The average measured amplitude is shown in Fig.~\ref{fig:pulse_amplitude_avg}. Note that the electrical output voltage of the SBD detector is proportional to the optical input power.

The MBI threshold can be identified slightly above \SI{0.2}{mA} (Fig.~\ref{fig:pulse_amplitude_avg}). Noteworthy is the different behavior of the diodes' signal at this threshold value. The forming of sub-structures above the MBI threshold only has a minor effect at the $D_1$ frequency band between \SIrange{90}{140}{GHz}. This is attributed to the already high amount of coherence from the bunch as a whole, so that the sub-structures increase mostly affects higher frequency components. The latter is evident in the data obtained in the two highest frequency bands. Without sub-structures, these diodes hardly detect any radiation while they are not sufficiently sensitive for incoherent radiation in this setup. Note that the sudden drop of detected power, seen for the two diodes $D_2$ and $D_3$ above \SI{0.4}{mA}, and for $D_2$, $D_3$ and $D_4$ around \SI{0.7}{mA} (cf.~Fig.~\ref{fig:pulse_amplitude_avg}) is due to a change of the bursting behavior. These sudden drops in the average values correlate with distinct features in the spectrograms (Fig.~\ref{fig:spectrogram_lin}), i.e. these average traces are highly reproducible and sudden drops are connected to the underlying electron density dynamics. For the simulation, $D_3$ and $D_4$ show a similar drop at about \SI{0.35}{mA}, but nothing similar around \SI{0.7}{mA}. There are two possible explanations for this small discrepancy. First, the simple parallel plates model does probably not fully reproduce the beam dynamics. Secondly, we assumed a flat spectral sensitivity for the simulation, while in the measurement the exact frequency of an emission peak is important (cf. Fig.~\ref{fig:sim_avg_spectrogram}).

If we assume a stable bunch shape, a quadratic increase of CSR with current is expected (see Fig.~\ref{fig:pulse_amplitude_avg}). This can be seen for the two lowest frequency bands, however, the bunch is not short enough for coherent radiation at the higher frequencies. Around the threshold, with the starting potential well distortion resulting in an asymmetric bunch shape and formation of substructures, the power at the higher frequencies increases stronger than quadratic as shown in our simulations (cf.~Figs.~\ref{fig:sim_avg_spectrogram} and~\ref{fig:sim_avg_spectrum}).

Above the threshold, the total emitted CSR power is stronger influenced by the changing form factor due to the sub-structures than by the number of the overall electrons. Subsequently, a larger power increase can be observed for a small current range at higher frequencies in comparison to the quadratic behavior (cf.~Fig.~\ref{fig:pulse_amplitude_avg}). This saturates quickly. Above \SI{0.7}{mA}, in the regime of the resonant bursts described before, the average power is determined by the length and repetition rate of the bursts. 

The lower panel in Fig.~\ref{fig:pulse_amplitude_avg} shows a similar behavior for the simulated data. In agreement with the measurements, the simulated and derived averages in the low frequency band ($D_1$) hardly show any change at the threshold, but similar changes are visible in the higher frequency bands. In the simulations, the threshold is found at a slightly higher bunch current. Considering that the only impedance taken into account is the CSR impedance shielded by parallel plates, \textit{Inovesa} reproduces the measured data very well. In conclusion, the measured phenomena can be attributed to the micro-bunching instability and can be simulated by a VFP solver.

% =====================
% Summary and Outlook
% =====================

\section{Summary and Outlook}
We have set up a single-shot 4-channel spectrometer system, which is able to continuously stream 500 million spectra per second. The optical setup consists of three beam splitters and four individual detectors, which are simultaneously read out in a streaming mode by the in-house build data acquisition system \textit{KAPTURE}. We have demonstrated the capability of the setup by synchronously observing the micro-bunching instability in four THz frequency bands turn-by-turn in a multi-bunch environment.

Many of the found features can be reproduced by simulations with our VFP solver \textit{Inovesa}. VFP solvers are known to determine the MBI threshold sufficiently well. Our comparison of the different diodes above the threshold region demonstrates that a simple model, considering only CSR shielded by parallel plates, does not only reproduce most statistical values, but also the bursting dynamics -- at least for KARA. The coincidence at all four frequency ranges is a very good indication that the simulated phase space fits to the real bunch, especially in the regular bursting regime where a clear phase offset between the different frequency ranges could be observed in measurement and simulation. This leads to new insights of the micro-bunching instability, where the missing parts of the measurements can be completed by the simulations.

Currently, we are extending \textit{Inovesa} from the basic parallel plates model to a more realistic impedance. The improved model includes resistive wall losses and will allow arbitrary impedances which could cover edge radiation with interaction in the subsequent straight sections to achieve an even better agreement between simulations and measurements. 

Developments are under way to use integrated single-chip detector arrays~\cite{YBCO-array1,YBCO-array2} based on superconducting, electric field-sensitive YBCO detectors~\cite{YBCO} and on room-temperature Schottky detectors~\cite{dresden_diodes} in combination with the next version of \textit{KAPTURE}. The array detectors will not only provide more narrow band responses but additionally increase the number of channels and reduce the required setup space.
The next KAPTURE version will provide eight readout channels and an improved readout path~\cite{KAPTUREc}. The ability to synchronize multiple KAPTURE boards would allow the readout of even more channels.

Moreover, synchronization with other experimental stations at KARA will additionally provide time-resolved information about the energy spread~\cite{kehrer} and the longitudinal bunch profile~\cite{eosd2}, simultaneously~\cite{kehrer-sync}, closing the circle of observing all projections of the longitudinal phase space.

\begin{acknowledgments}
We would like to thank Y.-L. Mathis and his team from the IBPT Infrared Group at KIT for support at the Beamlines. This work has been supported by the Initiative and Networking Fund of the Helmholtz Association under contract number VH-NG-320 and the German Federal Ministry of Education and Research (BMBF) under grant no. 05K16VKA. T.Boltz, M. Brosi, P. Sch{\"o}nfeldt and J. Steinmann acknowledge the support of the Helmholtz International Research School for Teratronics (HIRST) and E. Blomley the support of the Karlsruhe School of Elementary Particle and Astroparticle Physics (KSETA).
\end{acknowledgments}

\bibliographystyle{apsrev}
\bibliography{4DiodesReferences}
\end{document}